# Gate-Defined Graphene Quantum Point Contact in the Quantum Hall Regime

S. Nakaharai[1,2], J. R. Williams[1,3*] and C. M. Marcus[1]

1. Department of Physics, Harvard University, Cambridge, MA 02138
2. Toshiba Research and Development Center, Kawasaki 212- 8582, Japan
3. School of Engineering and Applied Science, Harvard University, Cambridge, MA 02138

We investigate transport in a gate-defined graphene quantum point contact in the quantum Hall regime. Edge states confined to the interface of p and n regions in the graphene sheet are controllably brought together from opposite sides of the sample and allowed to mix in this split-gate geometry. Among the expected quantum Hall features, an unexpected additional plateau at $0.5\, h/e^2$ is observed. We propose that chaotic mixing of edge channels gives rise to the extra plateau.

Graphene is a two-dimensional sheet of carbon atoms whose low-energy band structure is linear in momentum, resulting in unique transport properties [1-4]. One consequence of the linear dispersion is Landau quantization with $E \propto \pm \mathrm{Sqrt}(NB)$ ($N=0,1,2,\ldots$), where $B$ is the perpendicular magnetic field and $N$ is the Landau level (LL) index [5-6]. This, in turn, gives rise to novel quantum Hall (QH) features, including half-integer quantization of Hall conductance, $\sigma_{xy} = \pm 4(N+1/2)e^2/h$ ($N=0, 1, 2, \ldots$) [7-10].

Recent work on gated graphene devices has demonstrated local control of carrier type and density and the formation of p-n junctions [11-13]. Conductance across the p-n junction in the QH regime is fractionally quantized, a result of edge-channel mixing along the p-n junction, investigated theoretically in Ref. [14] as an instance of quantum chaotic scattering. Experiments verified the predicted fractional conductance values, $g = |v_p||v_n|/(|v_p|+|v_n|) \times e^2/h$ for p-n junctions [12], and $g = |v_p||v_n|/(|v_p|+2|v_n|) \times e^2/h$ for p-n-p junctions [13].

In this Letter, we extend the application of p-n junctions as a novel boundary in graphene by experimentally investigating a quantum point contact (QPC) geometry, defined by top gates that form voltage-controlled narrowly separated p-n interfaces (Fig 1). Longitudinal and Hall resistances were measured across the QPC in the QH regime, with quantizing magnetic fields applied perpendicular to the graphene surface over a range of temperatures from 0.25K to 20K. Gate-voltage-dependent resistances through the split-gate QPC were consistent with expected quantized values, but also yielded an unexpected extra plateau in longitudinal resistance at 0.5 $h/e^2$, with corresponding zero Hall resistance plateau. Alternative models leading to these extra plateaus, including lifting of spin or valley degeneracy, as well as chaotic mixing of edge states in the constriction, are considered.

Graphene flakes were prepared by mechanical exfoliation and deposited on a degenerately-doped Si wafer with a 285 nm surface thermal oxide [12]. Single layer graphene was identified by optical microscopy. Six electrical contacts of thermally evaporated Ti/Au (5/40 nm) were patterned by electron-beam lithography. A 30 nm $Al_2O_3$ gate insulator was then deposited by atomic layer deposition using a functionalization layer to promote adhesion [15]. Above the insulating layer, a pair of split gates with a gap width of 150 nm was then deposited, using the same methods as the contacts. A completed device is shown in Fig. 1(b), with an optical micrograph in Fig. 1(b). In Fig. 1(a), the two separated triangles denote top gates TG(L) and TG(R). Beneath the top gates are contacts C1 and C2, used to access electrically the top-gated regions. Note that C1 and C2 contact the graphene and are isolated from the top gates.

Electrical measurements [Fig. 1 (a)] used an ac current bias, $I_{ac}$, at 95 Hz, applied to contact D2 while contact S2 was grounded. Longitudinal resistance, $R_L = dV_L/dI_{ac}$, was based on voltage $V_L$ measured between D1 and S1. Hall resistance between the top-gated regions, $R_G = dV_G/dI_{ac}$, was measured between C1 and C2. Resistances were taken as a function of top-gate voltage ($V_{TG}$) and back-gate voltage ($V_{BG}$). The sample was measured in a $^3$He cryostat over a temperature range 0.25–20 K, with perpendicular fields up to 8T.

The filling factor in the top-gated region was influenced by both $V_{TG}$ and $V_{BG}$, while the region outside the top gate depended only on $V_{BG}$. This allows independent control of filling factors in both regions using combinations of $V_{TG}$ and $V_{BG}$. The conversion from back-gate voltage to density was obtained via a parallel-plate capacitor model [1]. The ratio of capacitances of the top-gate oxide and back-gate oxide was determined from the slope of the resistivity maximum in a plot of $R_L$ ($V_{TG}$, $V_{BG}$), following Ref. [9]. Conversion factors from gate voltage to density were used to determine the filling factor [1] in each region. Filling factors −2, 2, 6 and 10 for the top-gated regions, $v_{gate}$, as well as bulk region outside the gates, $v_{bulk}$, can be identified in Fig. 2(a). The charge neutrality point for both bulk (horizontal) and gate (inclined) regions are indicated in Fig. 2(a) by the solid lines between $v = 2$ and −2 regions.

Details of the highlighted parallelogram in Fig. 2(a) are shown in Fig. 2 (b), and configurations of edge channels of the parallelograms labeled with (c)-(h) are pictured in Figs. 2(c)-(h). When bulk and top-gated regions have the same filling factors [Fig. 2(d)], quantum Hall edge states run along the sample edge, uninterrupted, resulting in $R_L = 0$. In contrast, when filling factors in the bulk and top-gate regions differ [Figs. 2(c, e-h)], the current flows along edges defined by the top gates, with a constriction formed at the QPC. In this configuration, the constriction can be switched from open to pinched off, depending on gate voltages. Accordingly, several parallelograms in Fig. 2(b) are divided into open and pinched off regions, illustrated on the left side of Fig. 2(b). For the open constriction, edge channels pass through the QPC without backscattering and $R_L$ remains zero. The open state in each parallelogram corresponding to the region below the (red) dotted line for $v_{gate} > 0$, and above the line for $v_{gate} < 0$ in Fig. 2 (b). In a pinched off region, edge channels flow *across* the QPC, both from left to right and right to left, which induces a potential difference between S1 and D1, and a finite $R_L$. For the pinched off configurations, $R_L = [(|v_{bulk}|+|v_{gate}|)/|v_{bulk}||v_{gate}|] \, h/e^2$ [17], as shown in each parallelogram of Fig. 2(b).

A more detailed plot of $R_L$ ($V_{TG}$, $V_{BG}$) for $v_{bulk} = -2$ at $B = 5$ T is shown in Fig. 3(a). The value of $R_L$ for the case of a fully pinched off constriction is shown in each parallelogram (in unit of $h/e^2$) in Fig. 3(b) [13]. However, the potential profile created by the split gates can cause the filling factor in the QPC to take another value in the constriction region, $v_{QPC}$, as shown in Fig. 3 (a). Here, the boundary lines for $v_{QPC}$ are assumed from the slope of apparent stripes in $V_{TG}$-$V_{BG}$ mapping and the identical spacing as $v_{gate}$. Schematic Landau level (LL) energies as a function of position across the QPC, at values of $v_{gate}$ and $v_{QPC}$ corresponding to regions (d), (e) and (f) in Fig. 3(a), are illustrated in Figs. 3(d), 3(e) and 3(f), respectively. In the configuration shown in Fig. 3(d), $R_L$ is close to the theoretical value of $h/e^2$ [17], while the open constriction shown in Fig. 3(e) produces zero resistance through the QPC. In the configuration shown in Fig. 3(f), the top-gated region has filling factor +6, while $v_{QPC}$ is +2. Therefore only the $N$=0 LL contributes to backscattering. Note that in the configuration shown in region c in Fig. 3 (e), stripes parallel to the $v_{QPC}$ boundaries are observed. In fact, conductance across the constriction (from left gate to the right, measured in a two-terminal configuration between contacts C1 and C2) exhibited Coulomb blockade (not shown), presumably due to disorder-induced puddles in the QPC region [18] acting as quantum dots [19].

Figure 3(c) shows the comparison of experimental and theoretical $R_L$ values as a function of $V_{TG}$ for various values of $V_{BG}$. These values are indicated by color on the left side of Fig. 3(a). As $|V_{BG}|$ increases, $R_L$ agrees less and less well with the theoretical value of 1 $h/e^2$ for 0.25V< $V_{TG}$ <0.75V. In addition, when $|V_{BG}|$ exceeds 7.8 V, the increase of $R_L$ from 0 moves to larger $V_{TG}$ values. These phenomena indicate that the component of current going through the constriction increases as the $|V_{BG}|$ increases, which is consistent with the fact that the constriction tends to open as $|V_{BG}|$ is increased, thus requires larger $|V_{TG}|$ to close the constriction.

We next turn our attention to a QPC configuration just before the constriction is pinched off, that is, a state just between Fig. 3(d) and 3(e). Figures 4(a), 4(b) and 4(c) show the dependence of $R_L$ and $R_G$ on $V_{TG}$ for $V_{BG}$ in the range of $v_{bulk} = -2$ at temperatures $T = 1.4$ K, 4.2 K and 20 K, at a field of $B = 5$ T. In Fig. 4(a), $R_L$ shows a clear plateau at 0.5 $h/e^2$ around $V_{BG} = -7.6$ V for the range of $V_{TG}$ between 0.3 V and 0.6 V. The plateau is particularly evident at $T = 4.2$ K [Fig. 4(b)], and can be seen for fields up to 8T (data not shown). By $T = 20$ K, the plateau is washed out and barely visible [Fig. 4 (c)]. The value of $R_L$ on the plateau is half of the theoretical value of 1 $h/e^2$ for a pinched off QPC [17]. This means that half of the current is backscattered at the QPC and the other half is transmitted. Simultaneously, $R_G$ shows a plateau at 0 over the same $V_{TG}$ range. Note that $R_G$ changes its sign from negative to positive when the QPC is switched from open to pinched off; $R_G = 0$ therefore implies that the plateau corresponds to the state between open and closed, that is, the current is divided into two equal components resulting in the equal potential of the both sides of the top-gated regions, consistent with the $R_L$ behavior on the plateau.

Mechanisms that could lead to a plateau at $R_L = ½ h/e^2$ include Zeeman or valley splitting of the $N = 0$ LL with $\nu_{QPC} = 0$ state in the QPC, or chaotic mixing of edge modes in the QPC, similar to chaotic mixing along an extended p-n edge [14]. Zeeman splitting of LLs was observed in Ref. [20], though only at lower temperature and higher fields. Lifting of valley degeneracy by confinement could also produce a plateau at $R_L = ½ h/e^2$. There is an active debate on spin-first or valley-first and on the mechanism for splitting [21-24]. This may also be related to the increasing resistivity around the $N=0$ LL at large magnetic fields [25- 27]. Typically, splitting of LL requires higher fields than those used here (< 8T). We note, also, that the extra plateau at $R_L = ½ h/e^2$ does not become more pronounced at higher fields (not shown).

As an alternative, we propose robust chaotic mixing of edge channels in the constriction as the source of the extra plateau, comparable to the mechanism leading to the quenching of the Hall effect in ballistic Hall junctions [28]. As illustrated in Fig. 4 (f), the QPC in the QH regime has two incoming channels and two outgoing channels that move along p-n interfaces and serve as fully transmitting leads. In the constriction, currents in the two incoming channels are mixed, and lose memory of which channel from which they entered the region. Moreover, there is equal probability of leaving the QPC in either of the two outgoing channels. If this regime of chaotic scattering persists over an extended range of top-gate voltages, a plateau with characteristic half-backscatter will result.

Recent theory considers quantum chaotic mixing of counter-propagating current channels of $N=0$ LLs in a graphene p-n QPC [29]. The analysis is based on random hopping among the p-n puddles resulting from disorder, and yields behavior comparable to a chaotic quantum dot [30]. These results suggest that robust mode mixing occurs whenever the characteristic dwell time within the scattering region exceeds the time to diffuse through the QPC region and the dephasing time. Mesoscopic fluctuations in the junction are suppressed by elevated temperature, as in Ref. [28].

We thank Dima Abanin and Leonid Levitov for valuable discussion. We also thank Max Lemme, Ben Feldman and Ferdinand Kuemmeth for assistance in fabrication and measurement. SN acknowledges A. Kurobe, T. Tanamoto, A. Nishiyama and J. Koga for support.

*Present address: Department of Physics, Stanford University, Stanford CA 94305, USA.

**Figures**

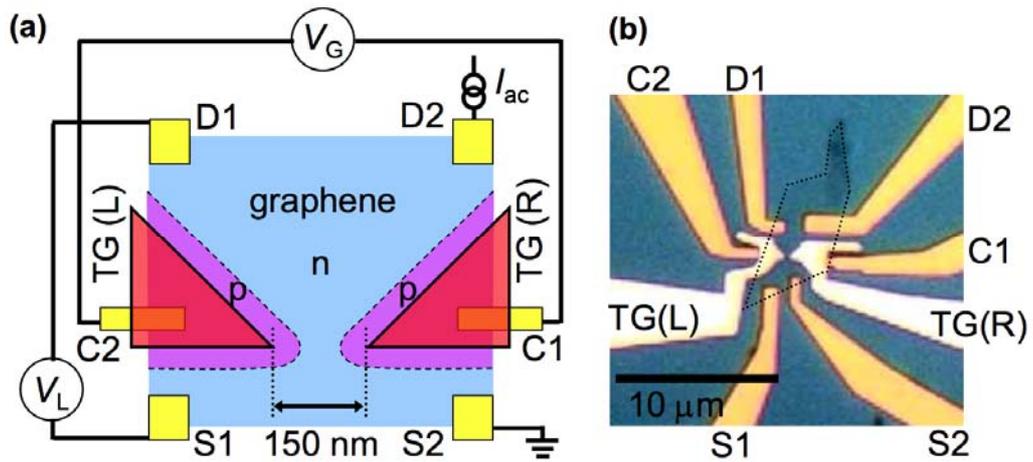

**Fig. 1.** (a) Schematic illustration of the graphene QPC device. Above the gate insulator, the split gates of TG(L) and TG(R) are formed with the designed gap width of 150 nm. The top-gate voltage $V_{TG}$, in conjunction with a global back-gate voltage $V_{BG}$ (not shown), controls carrier type (p or n) and density beneath TG(L) and TG(R). Carrier type and density outside the top-gated region is controlled solely by $V_{BG}$. Measurement configuration is also illustrated. (b) Optical micrograph of the device reported in the present work. Dotted lines represent the boundary of the graphene flake.

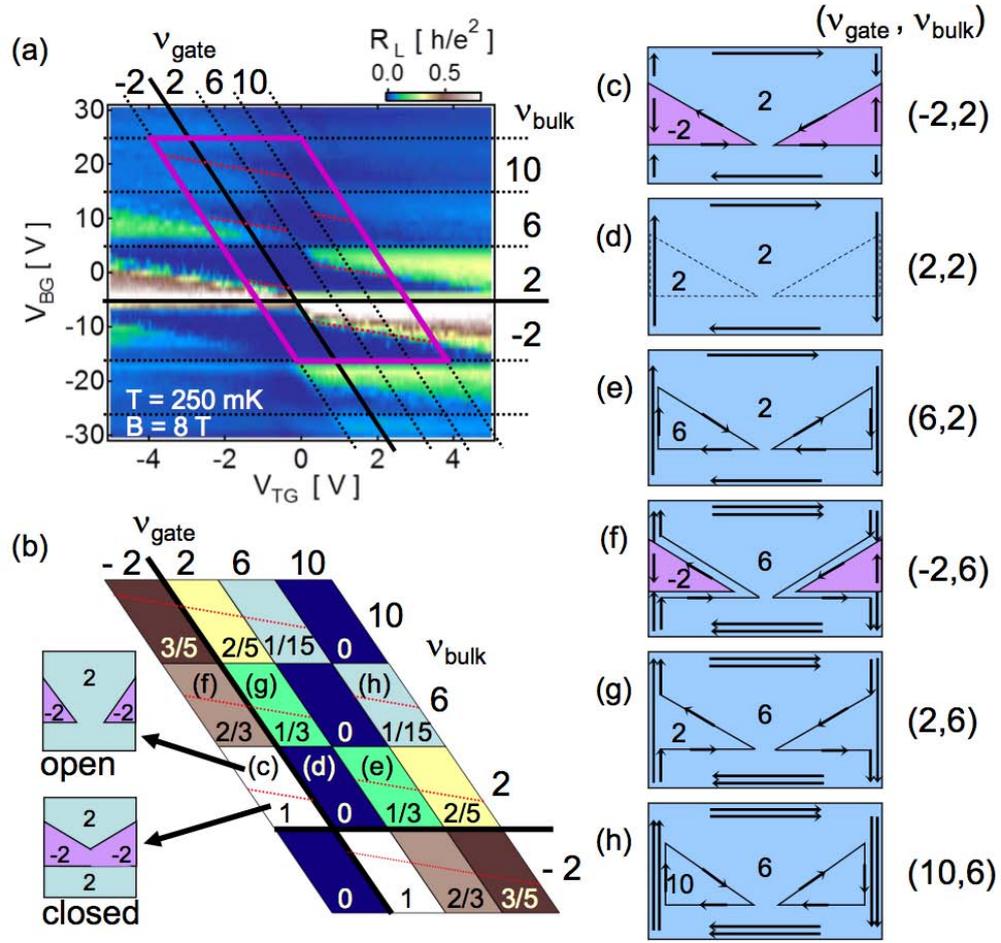

**Fig. 2.** (a) Longitudinal resistance, $R_L$, as a function of $V_{TG}$ and $V_{BG}$ field $B$ = 8 T and temperature $T$ = 0.25 K. Parallelograms mark regions of specific filling factors in the bulk ($\nu_{bulk}$) and top-gated ($\nu_{gate}$) region. The thick black lines represent the 0th LL for bulk (horizontal) and top-gated (inclined) regions, while dotted lines denote the higher LLs. (b) The details of the large parallelogram with a (purple) thick lines highlighted in (a). Numbers in each parallelogram indicate the theoretical value of $R_L$ in units of $h/e^2$ for a fully closed QPC. Each parallelogram is divided by a (red) dotted line as in (a) according to QPC open/closed configurations, as illustrated in the left of the ($\nu_{bulk}$, $\nu_{gate}$) = (2, −2). (c)-(h) Schematics of the current flow denoted by arrows for each parallelogram labeled as (c)-(h) in (b). The numbers correspond to the filling factors for the bulk and top-gated regions.

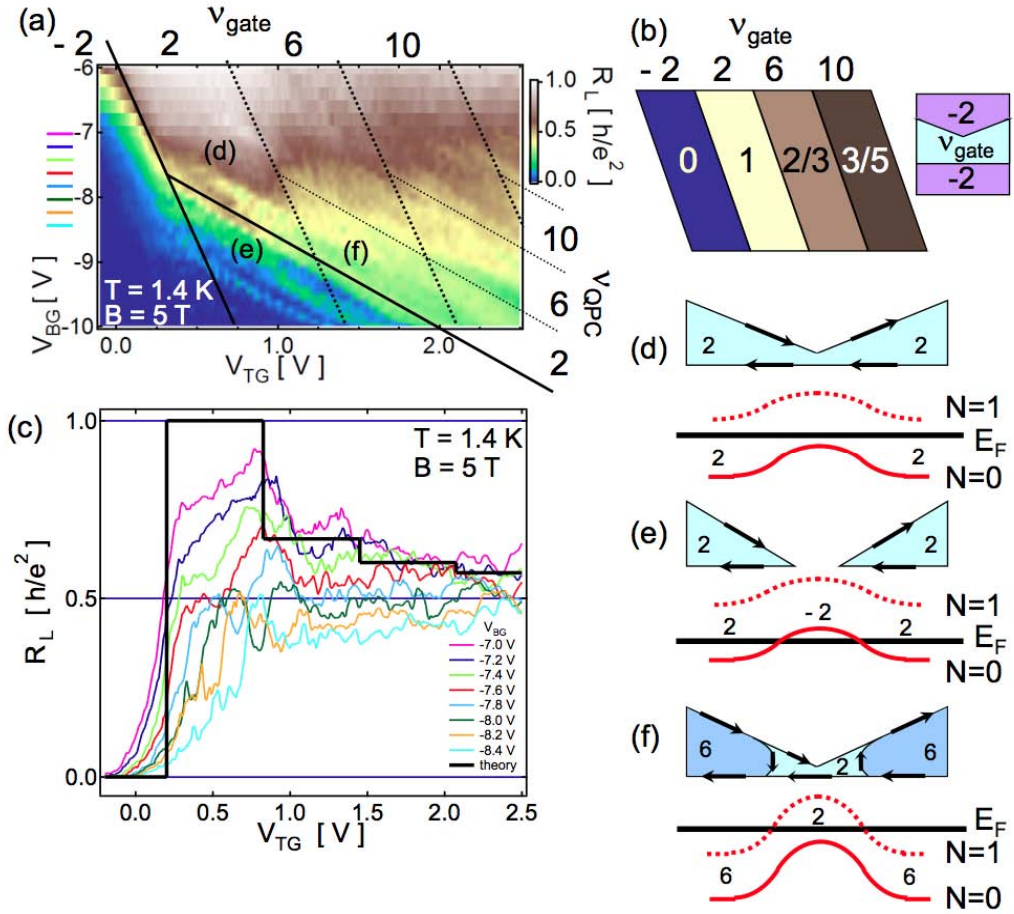

**Fig. 3.** (a) Longitudinal resistance, $R_L$, as a function of $V_{TG}$ and $V_{BG}$ at $B$ = 5 T and $T$ = 1.4 K with $\nu_{bulk} = -2$. (b) Theoretically expected $R_L$ values are given in a unit of $h/e^2$ for each parallelogram. (c) $R_L$ as a function of $V_{TG}$ for $V_{BG}$ between $-7$ V and $-8.4$ V at $T$ = 1.4 K and $B$ = 5 T. The theoretical $R_L$ value corresponding to the closed QPC at $V_{BG} = -7$ V (black thick line). Each curve has the corresponding marker on the left of (a) with the same color. (d), (e) and (f) show the shape of the top-gated region and the corresponding energy diagram across the QPC. Here, arrows represent the direction of the edge currents, $E_f$ is the Fermi energy, $N$ is the LL index and the numbers are the filling factors.

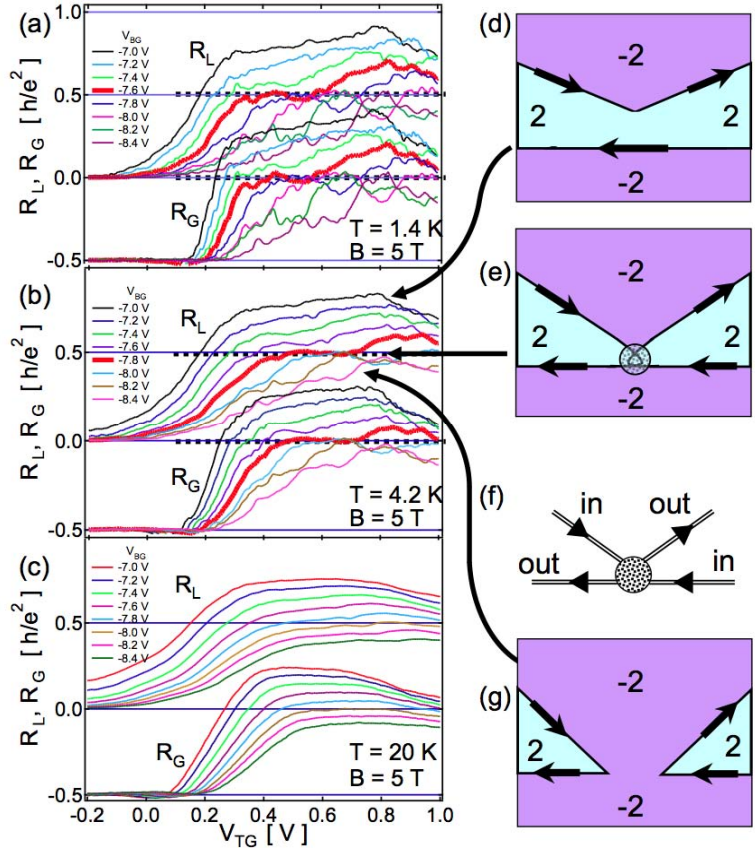

**Fig. 4.** Longitudinal resistance, $R_L$, and Hall resistance between top-gated regions, $R_G$, as a function of $V_{TG}$ with stepped $V_{BG}$ at $B = 5$ T and temperatures $T = 1.4$ K (a), 4.2 K (b) and 20 K (c). The state with completely closed QPC with $R_L > 0.5$ $h/e^2$ and $R_G > 0$ (d), while the open QPC with $R_L < 0.5$ $h/e^2$ and $R_G < 0$ is shown in (g). The state just between (d) and (g) is schematically shown in (e). (f) Conceptual diagram for chaotic mode-mixing configuration in the QPC in (e).